\documentclass{pasj00}

\begin{document}
\SetRunningHead{F. Nakata et al.}{Galaxy Population around the Radio Galaxy
3C 324 at $z=1.2$}
\Received{2001/05/28}
\Accepted{2001/09/25}

\title{Galaxy Population in a Cluster of Galaxies around the Radio
Galaxy 3C 324 at $z=1.2$\altaffilmark{1}}

\author{Fumiaki \textsc{Nakata},\altaffilmark{2}
        Masaru \textsc{Kajisawa},\altaffilmark{3}
        Toru \textsc{Yamada},\altaffilmark{4}
        Tadayuki \textsc{Kodama},\altaffilmark{2} \\
        Kazuhiro \textsc{Shimasaku},\altaffilmark{2}
        Ichi \textsc{Tanaka},\altaffilmark{4}
        Mamoru \textsc{Doi},\altaffilmark{5}
        Hisanori \textsc{Furusawa},\altaffilmark{2} \\
        Masaru \textsc{Hamabe},\altaffilmark{6}
        Masanori \textsc{Iye},\altaffilmark{4}
        Masahiko \textsc{Kimura},\altaffilmark{7}
        Yutaka \textsc{Komiyama},\altaffilmark{8} \\
        Satoshi \textsc{Miyazaki},\altaffilmark{4}
        Sadanori \textsc{Okamura},\altaffilmark{2}
        Masami \textsc{Ouchi},\altaffilmark{2}
        Toshiyuki \textsc{Sasaki},\altaffilmark{8} \\
        Maki \textsc{Sekiguchi},\altaffilmark{7}
        Masafumi \textsc{Yagi},\altaffilmark{4}
        and
        Naoki \textsc{Yasuda}\altaffilmark{4}
        }
\altaffiltext{2}{Department of Astronomy, School of Science, The University
of Tokyo, Bunkyo-ku, Tokyo 113-0033}\email{nakata@astron.s.u-tokyo.ac.jp}
\altaffiltext{3}{Astronomical Institute, Tohoku University, Aoba-ku, Sendai,
Miyagi 980-8578}
\altaffiltext{4}{National Astronomical Observatory of Japan, 2-21-1 Osawa,
Mitaka, Tokyo 181-8588}
\altaffiltext{5}{Institute of Astronomy, The University of Tokyo, Mitaka,
Tokyo 181-1500}
\altaffiltext{6}{Department of Mathematical and Physical Sciences, Faculty
of Science, \\ Japan Women's University, 2-8-1 Mejirodai, Bunkyo-ku, Tokyo
112-8681}
\altaffiltext{7}{Institute for Cosmic Ray Research, The University of Tokyo,
5-1-5 Kashiwa-no-Ha, Kashiwa, \\ Chiba 277-8582}
\altaffiltext{8}{Subaru Telescope, National Astronomical Observatory of Japan,
650 North Aohoku Place, Hilo, \\ HI 96720, U.S.A}


\maketitle

\footnotetext[1]{
Based in part on data collected at Subaru Telescope, which is
operated by the National Astronomical Observatory of Japan.
}

\begin{abstract}
We discuss the properties of galaxies around the radio galaxy 3C 324
at $z=1.2$ based on $BVRIK'$ multi-band imaging data. We have applied a
photometric-redshift technique to objects in the 3C 324 field, and identified
35 objects as plausible cluster members. We have found that red and luminous
members are concentrated in a small region enclosed by a circle of
40$''$ radius (0.33 Mpc at $z=1.2$ for $\Omega_0=0.3,$ $\lambda_0=0.7,$ $
H_0=70$~km~s$^{-1}$~Mpc$^{-1}$ cosmology) from the 3C 324 galaxy. The 3C 324
cluster is probably much more compact in size compared with the local
clusters. We constructed a $K'$-band luminosity function of the cluster
members and fit a Schechter function, and found the characteristic magnitude
to be $K'^\ast_{\rm AB}=20.2\pm0.6$. This value is consistent with the
extrapolation of the pure passive evolution seen for $z<1$ clusters. We
have identified eight bright galaxies which form a red color--magnitude
sequence. The slope of the sequence is consistent with the passive evolution
model down to $K'_{\rm AB}<22$; we also found that there is no clear age
variation in these bright red galaxies. However, seven out of these eight
galaxies exhibit a significant excess in the rest UV light with respect to
the passive evolution model. This may suggest that the massive early-type
galaxies in this high-redshift cluster are still forming stars to some extent.
We have confirmed a truncation of the color--magnitude sequence at $
K'_{\rm AB}\sim22$; faint passively-evolving galaxies may not yet be present
in this cluster at $z\sim1.2$. The overall color distribution of the cluster
members, selected by the photometric redshift technique, is found to be very
broad. We derived the fraction of blue galaxies in this cluster following a
definition of Butcher and Oemler (1984), and obtained $f_{\rm B}=0.39\pm0.28$,
which is higher than that for $z<1$ clusters. This indicates that the
star-formation activity of this cluster is, on the average, higher than that
of lower redshift counterparts.
\end{abstract}

\section{Introduction}

The early evolution of early-type (elliptical and S0) galaxies can provide a
strong constraint on galaxy-formation models. Because early type galaxies form
the dominant population in nearby rich clusters (Oemler 1974; Dressler 1980),
they are very efficiently studied in clusters. According to many important
observations made for clusters at $z\lesssim1$, it is well established that the
majority of the early-type galaxies observed in nearby and
intermediate-redshift ($z\sim0.5$) clusters mainly consist of old stellar

\newpage
\noindent
populations formed at high redshifts ($z_{\rm F}>2$, e.g., O'Connell 1988;
Bower et al.\ 1992; Arag\'{o}n-Salamanca et al.\ 1993; Rakos, Schombert
1995; Gladders et al.\ 1998; Ellis et al.\ 1997; Stanford et al.\ 1998;
Kodama et al.\ 1998).

On the other hand, there are also many pieces of evidence suggesting that the
evolution of the entire cluster galaxy population is not as simple as
expected from the monolithic galaxy formation scenario.
It is known that
the fraction of blue galaxies in clusters increases rapidly
with the redshift at $z\lesssim1$ (Butcher, Oemler 1978, 1984; Rakos, Schombert
1995). Through previous spectroscopic studies of intermediate-redshift
clusters, it is also known that a significant fraction of galaxies have
emission-lines and/or post-starburst signatures (e.g., Dressler, Gunn 1992;
Postman et al.\ 1998; Dressler et al.\ 1999; Poggianti et al.\ 1999).
Such blue and star-forming galaxies are notably rare in rich clusters at $
z\sim0$. Recent ground-based spectroscopic observation and Hubble Space
Telescope (HST) high-resolution imaging have further revealed a population of
old galaxies with ongoing or recent (within a few Gyr prior to the
observation) star-formation activity as well as a significantly large fraction
of late-type disk galaxies and closely interacting systems in intermediate
redshift clusters (Abraham et al.\ 1996; Morris et al.\ 1998; Oemler et
al.\ 1997; Couch et al.\ 1998; van Dokkum et al.\ 1998;
Poggianti et al.\ 1999). This star-forming activity of cluster galaxies is
possibly related to the evolution of the morphological mix of the cluster
galaxy populations, particularly, the origin of S0 galaxies (Dressler et
al.\ 1997; van Dokkum et al.\ 1998, 2000; Kuntschner, Davis 1998). As a natural
extension of these results, it is expected that a stronger
evolution of galaxies may be observed in clusters at higher redshifts.

Recently, more than several clusters and cluster candidates have been
discovered at $z\gtsim1$ (Dickinson 1995; Postman et al.\ 1996; Stanford et
al.\ 1997; Hall, Green 1998; Ostrander et al.\ 1998; Olsen et al.\ 1999;
Ben\'{\i}tez et al.\ 1999; Rosati et al.\ 1999; Tanaka et al.\ 2000;
Best 2000). By observing higher-redshift clusters, any difference in the
star-forming history can be seen more clearly. A rapid change in the colors
of passively evolving galaxies only occurs within $\sim$2--3~Gyr after the end
of the star formation (e.g., Bower et al.\ 1992, 1998). If cluster early-type
galaxies formed at $z>2$ and evolved passively, one can expect conspicuous
color changes only at $z\gtsim1$. One may also expect more frequent
star-formation activity at higher redshifts, since there are a significant
fraction of galaxies with poststarburst signatures in intermediate-redshift
clusters (Couch, Sharples 1987; Dressler, Gunn 1992; Barger et al.\ 1996;
Postman et al.\ 1998; Couch et al.\ 1998; Poggianti et al.\ 1999). A simple
extrapolation of the Butcher--Oemler effect (Butcher, Oemler 1984) also
predicts a blue galaxy fraction greater than 50\% at $z\sim1$. It is
therefore interesting to extend the detailed color analysis to other
clusters at $z\gtsim1$, and to probe the early history of star formation in
clusters.

To date, however, only a few clusters at $z\gtsim1$ have been studied
in detail by multicolor photometry (Stanford et al.\ 1997; Ben\'{\i}tez et
al.\ 1999; Rosati et al.\ 1999; Tanaka et al.\ 2000; Haines et al.\ 2001;
van Dokkum et al.\ 2001). The two largest difficulties in studying the galaxy
population in these high-redshift clusters are: (1) the faintness of the
cluster galaxies due to the large distance from us and (2) contamination
of the foreground and the background galaxies, as they tend to dominate the
galaxy surface number density on the sky towards the high redshift clusters.
We have overcome these difficulties by utilizing the huge light-collecting
power of the 8 m Subaru telescope and by applying a photometric-redshift
technique to isolate the possible cluster members.

We discuss in this paper the 3C 324 cluster at $z=1.2$. An excess of galaxy
surface number density in this region was recognized by Kristian, Sandage, and
Katem (1974) and by Spinrad and Djorgovski (1984), and firmly identified by
Dickinson (1995). In Dickinson's (1997b) spectroscopic study, the
surface-density excess was revealed to be due to the presence of two
clusters or rich groups at $z=1.15$ and $z=1.21$. Extended X-ray emission
with a luminosity comparable to that of the Coma cluster has been detected in
the direction of 3C 324 (Dickinson 1997a), which suggests that at least one of
the two systems is a collapsed massive system. From a deep HST image,
Smail and Dickinson (1995) detected a weak shear pattern in the field that may
be produced by a cluster. In the following discussion, we do not distinguish
the two clusters at $z\sim1.2$, since either a detailed redshift distribution
of the galaxies or the relative population of the two systems is not yet
available. We thus discuss the average properties of the two clusters.
Since their redshifts are close to each other, this has little effect on
our discussion about the luminosity and color distributions.

The 3C 324 cluster has been studied by Kajisawa et al.\ (2000a,b;
hereinafter referred to as K00a,b) with the HST $B_{\rm F450W}$- and $
R_{\rm F702W}$-bands and the Subaru $K'$-band data, which we include in this
analysis as well. K00a investigated the $K'$-band luminosity distribution of
galaxies around 3C 324. They found an abrupt decrease in the $K'$-band
luminosity function for galaxies in the cluster core region ($< 40''$ from
3C 324) at $K'_{\rm AB}\gtsim22$, and suggested two possible interpretations,
namely, luminosity segregation where a fraction of bright galaxies increases
toward the cluster center, or the intrinsic deficiency of faint galaxies in
the 3C 324 cluster. K00b also studied the optical and near-infrared color
of $K'$-selected galaxies in this cluster. They found a truncation of the
color--magnitude sequence of the 3C 324 cluster at $K'_{\rm AB}\sim22$.
They also argued that the bulge-dominated galaxies in the field seem to form
a broad sequence in the color--magnitude diagram, whose slope at the faint
end is much steeper than that expected from metallicity variations within a
passively evolving coeval galaxy population.

These results, however, are based on simple color selection or statistical
background subtraction of the field galaxies, and can be affected by a
contamination of the foreground/background objects. It is clearly needed to
test them using a sample of galaxies that are more firmly identified
as cluster members.

Spectroscopic redshifts are ideal, but not practical, at this high
redshifts ($z>1$), since even with 2 hours of integration on an 8 m-class
telescope, we can reach only down to $\sim M^\ast$. Therefore, it is extremely
time-consuming to make a complete sample over a representative field of
distant clusters (e.g., 1~Mpc). In this paper, we try to overcome this
difficulty in identifying cluster membership by using a photometric-redshift
technique. This approach enables us to investigate the galaxy properties of
likely cluster members; importantly, not only for the galaxies on the red
sequence, but also for relatively blue galaxies, which are otherwise embedded
in numerous field populations on a color--magnitude diagram. For this purpose,
we newly obtained $V$ and $I$ data with Suprime-Cam at the Cassegrain focus of
the Subaru telescope.

Firstly, we made a catalog of $K'$-band selected objects. The use of a $
K'$-band sample has the following two advantages. First, the selection bias at
large redshifts is small, and thus a direct comparison with lower-redshift
clusters is possible (Arag\'{o}n-Salamanca et al.\ 1993). Second, the sample
selection is less affected by the star-formation activity. We applied a
photometric-redshift technique by utilizing multi-band imaging data, and
identified 35 objects as plausible cluster members. With this new cluster
membership information as well as two additional colors essential for a
population analysis, we revisited the issues which remain unsettled in K00a,b.
With the photometric-redshift selected sample, we showed that an abrupt
decrease in the faint end of the $K'$-band luminosity function in the cluster
core region shown in K00a, which is mainly because the fraction of faint
members relative to the bright members is lower in the core region than in
the whole cluster, i.e., luminosity segregation; we also found that K00a
regarded some faint members as non-members. The truncation of the
color--magnitude sequence, indicated in K00b, was clearly identified in
this work as well. However, the broad and tilted color--magnitude sequence
argued in K00b is not completely recovered by the photometric-redshift
selected sample used in this paper. In addition to these interesting results,
we newly found a UV-excess in some early-type galaxies, and also derived
the blue fraction of the 3C 324 cluster at $z\sim1.2$.

This paper is structured as follows. We describe the observations and data
reduction briefly in section 2. Section 3 describes the object detection and
number counts. In section 4, we describe the photometric-redshift technique
and its application to our data. In section 5, we consider the spatial
distribution, luminosity function, color properties and the blue galaxy
fraction of the member galaxies in the 3C 324 cluster. We summarize the main
results in section 6.

Throughout this paper we use the following cosmological parameters: $H_0=70
$~km~s$^{-1}$~Mpc$^{-1}$, $\Omega_0=0.3$ and $\lambda_0=0.7$, which gives a
physical scale of 8.30~kpc~arcsec$^{-1}$ at the cluster redshift.

%
%

\begin{figure*}
   \begin{center}
      \FigureFile(170mm,170mm){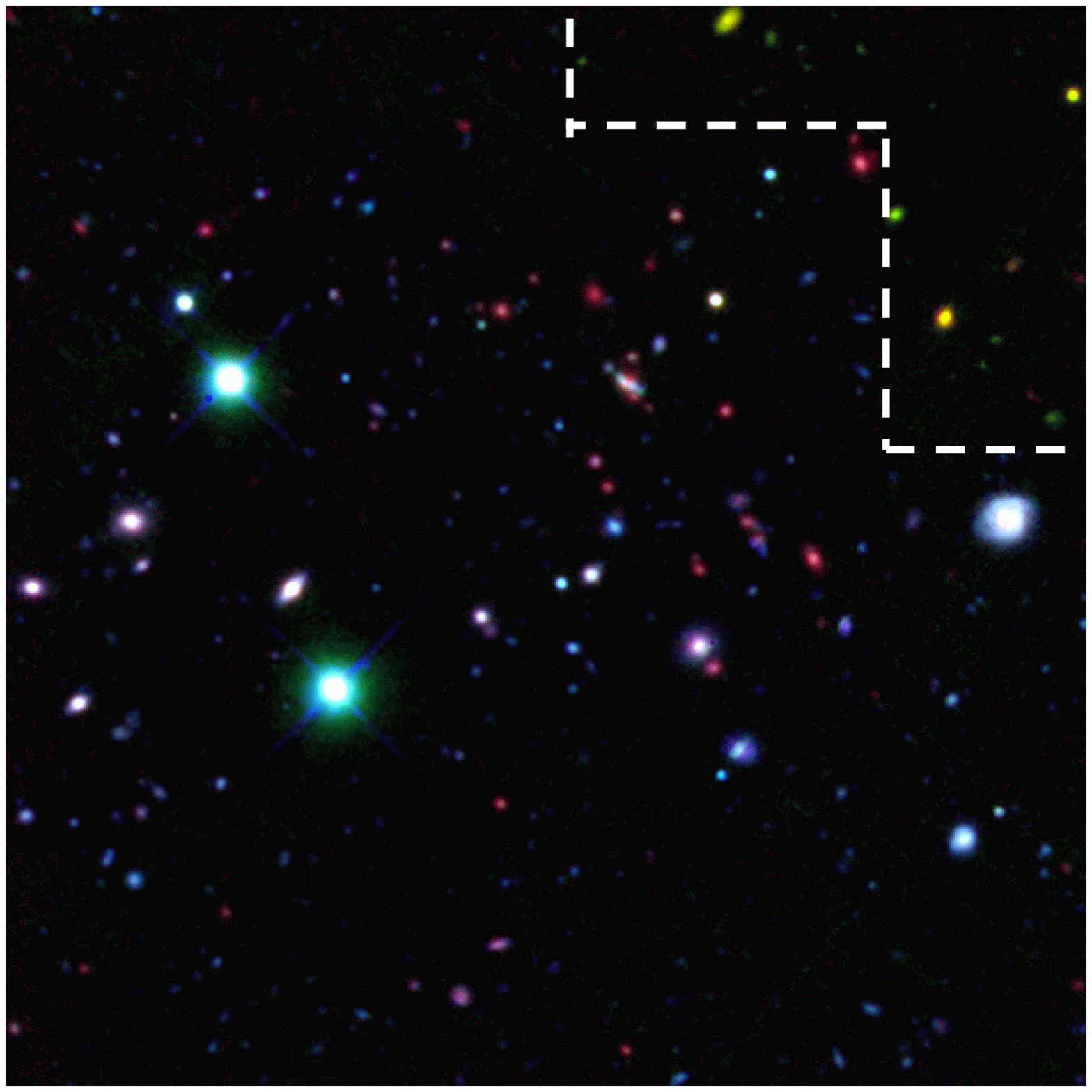}
   \end{center}
   \caption{Composite pseudo-color picture of the observed 3C 324 field.
   The RGB colors are assigned for $K',$ $I,$ and $R_{\rm F702W}$ images,
   respectively. The dashed line outlines the field of the HST observation;
   the upper-right region of the short-dashed line has not been observed with
   the HST.\label{fig1}}
\end{figure*}

\section{Data}

The $K'$-band and HST $B_{\rm F450W}$- and $R_{\rm F702W}$-band images used
in this paper were taken from K00a,b and Kajisawa and Yamada (1999). We newly
took $V$ and $I$ data with the Subaru telescope, which are described in
detail below.

\subsection{Suprime-Cam $V$- and $I$-band Data}

Suprime-Cam (Miyazaki et al.\ 1998) is a mosaic CCD camera designed for use
at the prime focus of the Subaru telescope with a field of view (FOV) of 30$'
\phi$. It is equipped with three-edge buttable 2k-by-4k CCDs in a $2\times5
$ array, giving a total of 8k$\times$10k effective pixels with a
pixel size of 15 $\mu$m. Suprime-Cam covers an FOV of $24'\times30'$, almost
the entire FOV of the F/1.86 prime focus, with a resolution of
0.\hspace{-2pt}$''$2 per pixel.

The 3C 324 field was observed in the $V$- and $I$-bands with a Suprime-Cam
mounted on the Cassegrain focus of the Subaru telescope during test observation
runs for commissioning of the Cassegrain focus in May--June 1999. The FOV
of the F/12.2 Cassegrain focus is 6$'$ in diameter; its physical size is
almost the same as that of the prime focus. Only six of the 10 CCDs, however,
were ready at the time of the $V$ and $I$ observations. Accordingly,
Suprime-Cam at the Cassegrain focus covered an FOV of $4'\times 3'$ with a
resolution of 0.\hspace{-2pt}$''$03 per pixel.

A log of the observations is given in table 1. The total integration times
of $V$- and $I$-bands are 2 hr and 1.5 hr, respectively. The seeings of
the combined image are $\sim0.\hspace{-2pt}''7$ for $V$ and $\sim
0.\hspace{-2pt}$''$6$ for $I$. The weather condition was stable during the $V
$-band observation. The $I$-band images, however, were taken under poorer
conditions. The sky was not photometric during the $I$-band observation.

%
%

\begin{table}
\begin{center}
\caption{Log of the observations.}
 \begin{tabular}{lcccc}
  \hline\hline
  Date & Band & Exposure & Frames & Seeing \\ \hline
  18/5/99 & $I$ & 15 min & 4 & 0.\hspace{-2pt}$''$5 \\
  19/5/99 & $I$ & 15 min & 2 & 0.\hspace{-2pt}$''$6 \\
  11/6/99 & $V$ & 15 min & 4 & 0.\hspace{-2pt}$''$7 \\
  12/6/99 & $V$ & 15 min & 4 & 0.\hspace{-2pt}$''$7 \\
  \hline
 \end{tabular}
\end{center}
\end{table}

Data reduction of each frame in a given band proceeded as follows. First, a
bias value was determined for each row of a frame as the median of the
overscan region. By subtracting the bias value from each row, an
overscan-subtracted frame was made. Then, the median bias frame was made by
taking the median of the available overscan-subtracted bias frames. This median
bias frame, whose average DC level is zero, was subtracted from all of the
overscan-subtracted frames so that the residual bias pattern would be
eliminated. Since the temperature of the dewar of Suprime-Cam was not
sufficiently low ($\sim-75^\circ$C) and was unstable during the observations,
the noise level due to dark currents was high and varied. Thus, a dark frame
was created for each of the object frames, depending on the temperature of
the dewar at the time when the frame was exposed.

Next, flatfielding was performed. Because the $V$-band images suffered from
stray light, the flat-field frame for the $V$-band was made from twilight sky
frames in which the counts of stray light were negligible. For the $I$-band,
however, the flat-field frame was made from dome-flat frames, because
twilight sky frames showed a strong fringe pattern. All of the frames were
flat-fielded. Thus, the fringe frame was made by taking the median of the
object frames; it was then subtracted from each of the flatfielded object
frames. Since the dome flat may not reflect the global pattern of the CCD
sensitivity, we further corrected the $I$-band frame for the global
sensitivity pattern. The sky flat frame was made by taking the median of the
fringe-subtracted object frames. All of the object frames were flat-fielded
again using this sky flat frame. The counts on those pixels contaminated with
cosmic rays were replaced with values interpolated from the surroundings.
Bad pixels were replaced with blanks, which were ignored in the following
analysis.

After subtracting the local sky background level, defined by taking the median
of the pixel values of the surrounding pixels, registering the relative
positions of the dithered frames, and normalizing the sensitivity of the six
chips, we finally co-added the individual exposures. We took the median in
this process. We applied $3\times3$ binning to obtain the final images in
order to avoid over-sampling. The actual resolution of the final images
is therefore 0.\hspace{-2pt}$''$09 per pixel.

For the $V$-band image, a flux calibration was performed using the
Landolt (1992) standard stars taken during the observation.
For the $I$-band image, which was not taken under the photometric condition,
we used the HST image in the F791W band (exposure time of 1800 s, PI: M.
Longair; PID: 1070) for the calibration. We acquired this F791W image from
the HST Data Archive; the zero point was kindly provided by Best (2000).
Although the transmission curves of the F791W filter and our $I$-band filter
are slightly different, the color term was found to be negligible. We used 16
common objects in these two images to determine the zero-point of our $
I$-band image. The standard deviation of F791W$-I$ colors of the 16 objects
around the average F791W$-I$ value was only 0.037 mag.

\subsection{CISCO $K'$-band Data}

$K'$-band imaging of 3C 324 field was carried out with a near-infrared camera,
CISCO (Motohara et al.\ 1998), mounted on the Cassegrain focus of the Subaru
Telescope on 1999 March 31 and April 1, during the telescope commissioning
period. The detector was a $1024\times1024$ HAWAII HgCdTe array with a pixel
scale of $0.\hspace{-2pt}$''$116$, which provided a field of view of $\sim2'
\times2'$. Details concerning the reduction procedures are described in
K00a. A flux calibration was performed by the UKIRT Faint Standard FS 27.
The seeing size of the resultant frame was $0.\hspace{-2pt}''8$, which was
measured from the FWHM of the stellar images.

\subsection{HST $B_{F450W}$ and $R_{F702W}$ Data}

The 3C 324 field was observed using the WFPC2 camera with a $R$ (F702W) filter
and a $B$ (F450W) filter (PI: M. Dickinson; PIDs 5465 and
6553, respectively). These data were taken from the HST Data Archive. The
total exposure times were 64800 s for the F702W image and 17300 s for the
F450W image.

\section{$K'$-band Selected Sample}

\subsection{Object Detection and Photometry}

We registered the $B_{\rm F450W},$ $V,$ $R_{\rm F702W},$ and $I$ images with
the $K'$ image, and applied Gaussian filters to smooth all of the images
to $\sim0.\hspace{-2pt}"8$ resolution of the $K'$ image. Only the region imaged
with the CISCO is considered in the following analyses. A composite
pseudo-color picture of the 3C 324 field is shown in figure 1, where the $R
$, $G,$ and $B$ colors are assigned for the stacked $K',$ $I,$ and $
R_{\rm F702W}$ images, respectively. The dashed line outlines the field of
the HST observation: the upper right region of the dashed line has not been
observed with the HST.

%
%

\begin{figure}
   \begin{center}
      \FigureFile(85mm,85mm){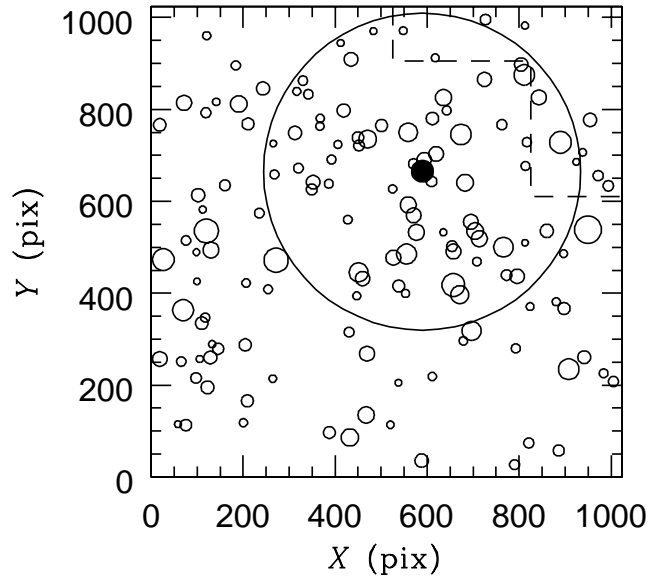}
   \end{center}
   \caption{Distribution of the 139 detected objects on the sky. 3C 324 is
   shown as the filled circle. The areas of the circles are in proportion to
   the brightness of objects. The dashed line outlines the field of the HST
   observation, similar to that shown in figure 1. The large circle indicates
   the `inner' region within 40$''$ from 3C 324.\label{fig2}}
\end{figure}

Since the $K'$-band, whose effective wavelength is longer than 4000\AA\ in
the rest frame at $z\sim1.2$, is essential for our photometric redshift
analysis (Kodama et al.\ 1999), we first constructed a catalog of $K'$-band
selected objects, and then searched for their counterparts on other optical
frames. Object detection on the $K'$-band frame was carried out by the
so-called `connected pixel method'. We defined an `object' as a clump of
more than $N_{\rm min}$ connected pixels with counts larger than an isophotal
threshold $I_{\rm thres}$. In this study, $I_{\rm thres}$ was set to 23.50
mag arcsec$^{-2}$, which is $\simeq1.12$-times the sky r.m.s. noise of
the $K'$-band image; $N_{\rm min}$ was set to 40 pixels (0.54 arcsec$^2$).
Photometry was performed with an aperture having a radius of 2$r_{\rm Kron}$,
where $r_{\rm Kron}$ is the Kron radius (See Metcalfe et al.\ 1991 for the
definition) of the object. We did not apply any star/galaxy separation so as
to avoid any possible omission of compact galaxies common at high
redshifts; however, the star contamination is expected to be negligibly small
at the faint magnitudes ($K'_{\rm AB}\gtsim19.5$) where we are mainly
interested. A total of 139 sources were detected at magnitudes brighter
than $K'_{\rm AB}=24.0$, which is the 3$\sigma$ detection limit of
the $K'$ image. Figure 2 shows the distribution of the detected objects on
the sky. The areas of the circles are in proportion to the brightness
of the detected objects. The filled circle indicates the position of the
radio galaxy 3C 324. The large-circle shows a region of 40$''$ radius centered
on 3C 324, which is discussed in subsection 5.1. The dashed line outlines the
field of the HST observation, similar to that shown in figure 1.

\subsection{Number Counts}

%
%

\begin{figure}
   \begin{center}
      \FigureFile(85mm,85mm){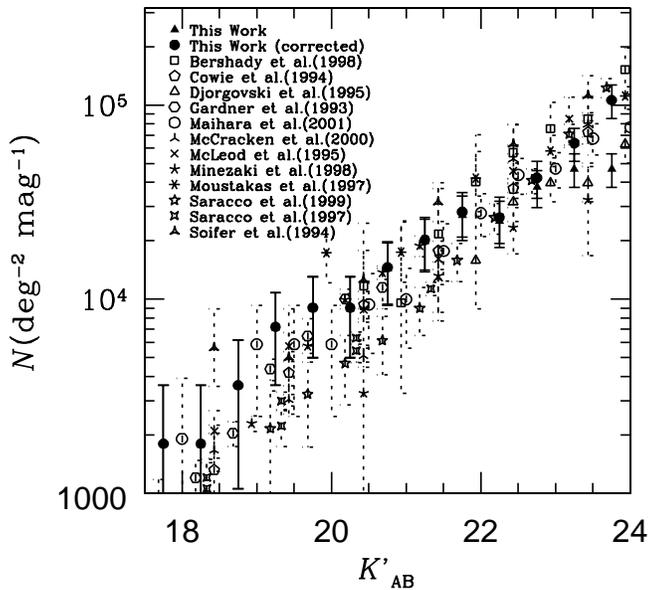}
   \end{center}
   \caption{$K'$-band number counts of the galaxies in the 3C 324 region.
   Filled triangles show raw counts, and filled circles indicate the counts
   corrected for incompleteness. Galaxy counts in the general field taken
   from various literature are also plotted using different symbols.
   \label{fig3}}
\end{figure}

The differential number counts of galaxies detected on the entire field of
the CISCO $K'$-band image are shown in figure 3. For a comparison, counts
obtained in general fields taken from various literature are also shown. The
filled triangles show our raw counts, and the filled circles indicate those
counts corrected for incompleteness. The completeness of object detection as a
function of the magnitude was adopted from K00a, who estimated the copleteness
factor at a given magnitude, using the IRAF ARTDATA package, by distributing
a number of artificial objects of that magnitude on the observed frame,
running the object detection software on the frame in the same manner as
for real objects, and deriving the detection rate of the artificial objects.

The corrected counts of the 3C 324 field are consistent with those of the
general fields within the error bars. Note, however, that the 3C 324 counts
are slightly higher than the average of the general fields at $19<K'_{\rm AB}
<22$, which probably reflects the existence of the 3C 324 cluster. At $
K'_{\rm AB}>22$, however, no apparent excess is seen. This abrupt decrease in
the number count at the faint end may be due to a field variation, which is
quite plausible for a small field coverage only $2'\times2'$ for the $
K'$-image.

\section{Photometric Redshift}

To extract members of the 3C 324 cluster at $z\sim1.2$, we applied a
photometric-redshift technique to those objects with $K'_{\rm AB}<24$, which
corresponds to the 3$\sigma$ detection limit of the $K'$-band image.
We estimated the redshifts of these object using the HYPERZ code of Bolzonella,
Miralles, and Pell\'{o} (2000). HYPERZ uses Bruzual and Charlot's (1993)
stellar evolutionary code (GISSEL98) to build synthetic template galaxies
with 8 star-formation histories, which roughly match the observed properties
of local galaxies from E to Im types: an instantaneous burst, a constant
star-forming system, and six exponentially decaying SFRs with time-scales of
from 1 to 30~Gyr. These models assume solar metallicity and a Millar--Scalo
IMF, and internal reddening is considered using the Calzetti et
al.\ (2000) model with $A_V$ varying between 0 and 1.2 mag.

We used $B_{\rm F450W},$ $V,$ $R_{\rm F702W},$ $I$ and $K'$ bands to estimate
the photometric redshifts, which covers from 2060\AA\ ($B_{\rm F450W}$) to
9721\AA\ ($K'$) in the rest-frame effective wavelengths at a cluster
redshift of 1.2. Because they properly bracket the 4000\AA\ break with a
sufficient number of bands, the photometric redshift estimation is expected to
achieve a high performance (e.g., Kodama et al.\ 1999). However, since the
colors of galaxies at $z=1.2$ in the $R$-, $I$-, and $K'$-bands are similar to
those at the $z>2$, some cluster members can be misidentified as $z>2
$ galaxies. Therefore, we limit the range of the output redshifts from HYPERZ
to $0\leq z\leq2$. That is, HYPERZ searches for the best-fit photometric
redshifts of individual galaxies in the range $0\leq z\leq2$.

%
%

\begin{figure}
   \begin{center}
      \FigureFile(85mm,85mm){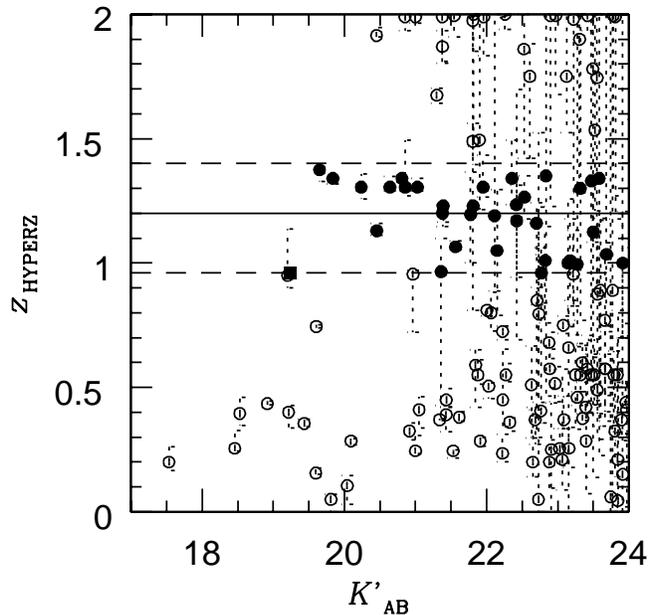}
   \end{center}
   \caption{Estimated photometric redshifts as a function of the $K'_{\rm AB}
   $ magnitude. The filled square shows 3C 324, and the filled circles
   indicate the cluster members, whose estimated photometric redshifts fall
   in the range $z=0.96$--1.4. The error bars indicate the 1$\sigma$ errors,
   which are taken from outputs of the HYPERZ code. \label{fig4}}
\end{figure}

Figure 4 shows the estimated photometric redshifts as a function of the $
K'_{\rm AB}$ magnitude. The filled square shows 3C 324. The error bars
indicate the 1$\sigma$ errors, which are taken from outputs of the HYPERZ
code. We find from figure 4 that $\simeq15$ galaxies with $K'_{\rm AB}
\ltsim22$ make a sequence around $z\sim1.2$, which is near to the redshift of
the 3C 324 cluster. Almost all of those galaxies, including 3C 324 (filled
square), are distributed at $0.96\leq z\leq1.4$. Therefore, we decided to
classify those galaxies whose estimated redshift is $0.96\leq z\leq1.4$ as
cluster galaxies. The lower cut, $z=0.96$, was chosen in order not to include
the very bright ($K'_{\rm AB}=19.2$) but blue ($I_{\rm AB}-K'_{\rm AB}=1.42
$) galaxy with an estimated redshift of 0.95, which is likely to be a
foreground galaxy. Thirty-five out of the total 139 galaxies fall in the
range of $z=0.96$--1.4 (filled circles). We regard those 35 galaxies as
members of the 3C 324 cluster (hereafter cluster members). Note that even if
we change the boundaries by $\Delta z\sim0.02$, the total number of cluster
members changes by at most 3, which does not significantly affect our results
given in the next section. We find from figure 4 that the estimated photometric
redshifts of the brightest members tend to concentrate at $z=1.3$--1.4,
slightly higher than the spectroscopic redshift of 3C 324 ($z=1.21$) and some
surrounding galaxies ($z=1.15,$ $1.21$; Dickinson 1997b). This difference
probably comes from a mismatch between the model SEDs and the observed
colors; $\Delta(I_{\rm AB}-K'_{\rm AB})\sim0.1$ corresponds
to $\Delta z\sim0.1$ (See figure 11).

%
%

\begin{figure}
   \begin{center}
      \FigureFile(85mm,85mm){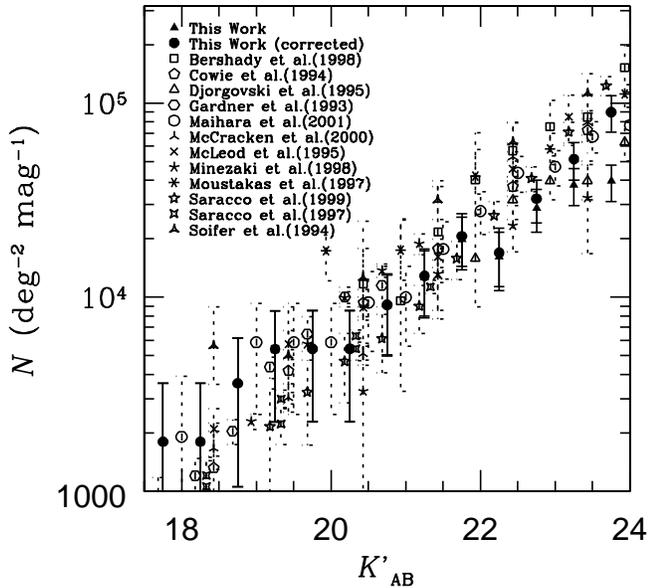}
   \end{center}
   \caption{$K'$-band number counts of galaxies in the 3C 324 field after
            removing the 35 cluster members. The symbols are the same as in
            figure 3. \label{fig5}}
\end{figure}

Figure 5 shows the differential number counts of galaxies in the 3C 324 field
after removing the 35 cluster members. The symbols are the same as in figure
3. The counts corrected for incompleteness are fairly consistent with those of
the general fields at $K'_{\rm AB}<22$. Note, however, that the 3C 324 counts
at $K'_{\rm AB}>22$ are lower than the average of general fields. These may be
due to a field variation, as discussed in subsection 3.2.

We checked the stability of our photometric redshift derived from the HYPERZ
package against the photometric errors as follows. First, we made a mock
catalog of 1000 galaxies for each of the magnitude bins of $\Delta m=0.5$ mag
in $K'_{\rm AB}=18$--24 mag at $z=1.2$ using a code provided in
HYPERZ for the same bandpasses and limiting magnitudes as for the 3C 324
data. The generated galaxies are among the model `template' galaxies used for
estimating the photometric redshifts, and were randomly adopted from the 8
models (E to Im). We then applied HYPERZ to those galaxies to estimate the
photometric redshifts. We found that about 80\% of the galaxies with $
K'_{\rm AB}<23$ fall in the range $0.96\leq z\leq1.4$, implying that the
criterion for membership adopted in this paper is appropriate. We found,
however, that only $\simeq50$ \% of the faint ($23\leq K'_{\rm AB}\leq24
$) galaxies satisfy this criterion, because of large photometric errors.
Therefore, we treat galaxies with $K'_{\rm AB}>23$ carefully in the following
sections. We also checked the fraction of $z>2$ galaxies with the catalog of
the Hubble Deep Field (HDF; Williams et al.\ 1996), whose redshifts were
estimated with a photometric-redshift technique by Fern\'{a}ndez-Soto,
Lanzetta, and Yahil (1999). We found that the fraction is about 6\% at $
K'_{\rm AB}<23$, and about 13\% at $K'_{\rm AB}<24$. Therefore, the
contamination of $z>2$ galaxies is negligible at $K'_{\rm AB}<23$. However,
at $23<K'_{\rm AB}<24$ the contamination can be a problem. It is worth noting
that the blue galaxy fraction of cluster members, which we discuss in
subsection 5.4, was derived from galaxies at $K'_{\rm AB}<23$, and thus the
uncertainty in the photometric-redshift technique is not very large.

%
%

\begin{figure}
   \begin{center}
      \FigureFile(85mm,85mm){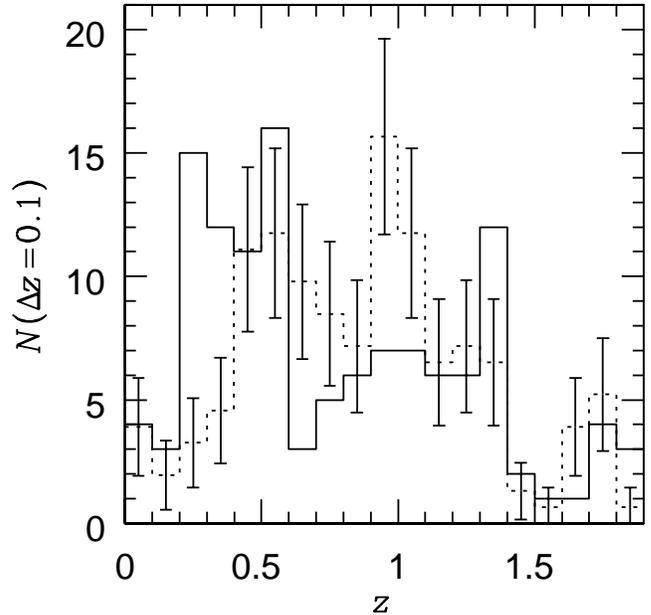}
   \end{center}
   \caption{Distribution of the estimated photometric redshifts for the $
   K'_{\rm AB}<24$ galaxies (solid line). For a comparison, the observed
   redshift distribution of field galaxies with $K'_{\rm AB}<24$ in the HDF
   is also plotted (dashed line) with the 1$\sigma$ Poisson errors. The
   redshift distribution of the HDF is normalized so that the number of
   galaxies is the same as of our 3C 324 field. \label{fig6}}
\end{figure}

To estimate the field contamination, we compared the redshift distribution of
the 3C 324 field with those of the HDF. Figure 6 shows the estimated
photometric redshifts (solid line) for the $K'_{\rm AB}<24$ galaxies. As a
reference, the observed redshift distribution of field galaxies with $
K'_{\rm AB}<24$ in the HDF is also plotted (dashed line) with the 1$\sigma
$ Poisson errors. Note that most of the redshifts of the HDF galaxies were
estimated by the photometric-redshift technique, but with a different
code (Fern\'{a}ndez-Soto et al.\ 1999). The redshift distribution of the HDF
was normalized so that the number of galaxies would be the same as that of
our 3C 324 field. We find from figure 6 that the field contamination expected
from the HDF at $0.96<z<1.4$ is significant. However, we should note that the
galaxies in the 3C 324 field which are assigned photometric redshifts in the
range $0.96<z<1.4$ mostly have $20<K'_{\rm AB}<24$. And at this magnitude
range, the number counts of the field galaxies toward the 3C 324 cluster are
systematically lower than the average, as shown in figure 5, possibly by up
to a factor of 3. Therefore, we cannot estimate the field contamination
precisely.

\section{Results}

\subsection{Distribution of Member Galaxies}

%
%

\begin{figure}
   \begin{center}
      \FigureFile(85mm,85mm){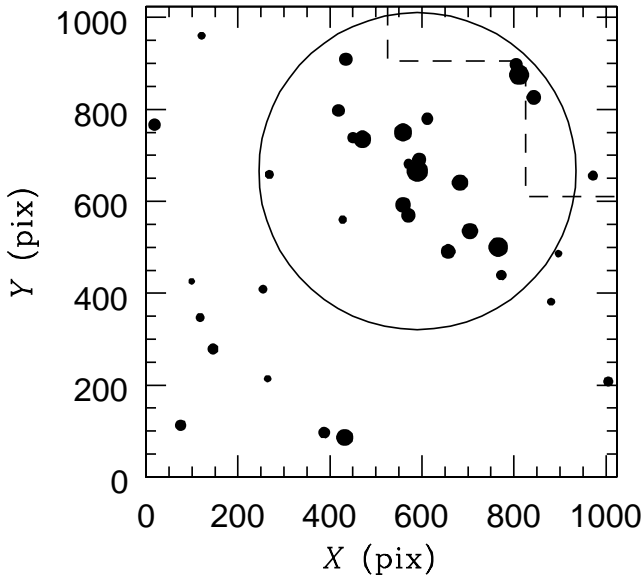}
   \end{center}
   \caption{Distribution of the 35 cluster members on the sky. The areas of
   filled circles are in proportion to the brightness of galaxies. The dashed
   line shows the boundary of the field of the HST observation. The large
   circle indicates the `inner region'. \label{fig7}}
\end{figure}

Figure 7 shows the distribution of the cluster members on the sky. The areas
of the filled circles are in proportion to the brightness of the galaxies. The
dashed line shows the boundary of the field of the HST observation.
For the two galaxies outside the HST field, we used only $V,$ $ I,$ and
$K'$ data to estimate their redshifts. We divided the observed field into the
following two regions: the `inner' region, which is enclosed by a circle of
40$''$ radius centered at 3C 324 (the large circle in figure 7 and figure 2),
and the adjacent `outer' region, which is the remaining region. A radius of
40$''$ corresponds to 0.33~Mpc at $z=1.2$ for $\Omega_0=0.3,$ $\lambda_0=0.7$,
$h=0.7$ cosmology. The areas of the inner and outer regions are 1.40
arcmin$^2$ and 2.60 arcmin$^2$, respectively. A total of 21 galaxies among
the 35 cluster members are in the inner region and 14 are in the outer
region. Therefore, the number density of cluster galaxies is 15.0 arcmin$^{-2}
$ and 5.4 arcmin$^{-2}$, respectively.

%
%

\begin{figure}
   \begin{center}
      \FigureFile(85mm,85mm){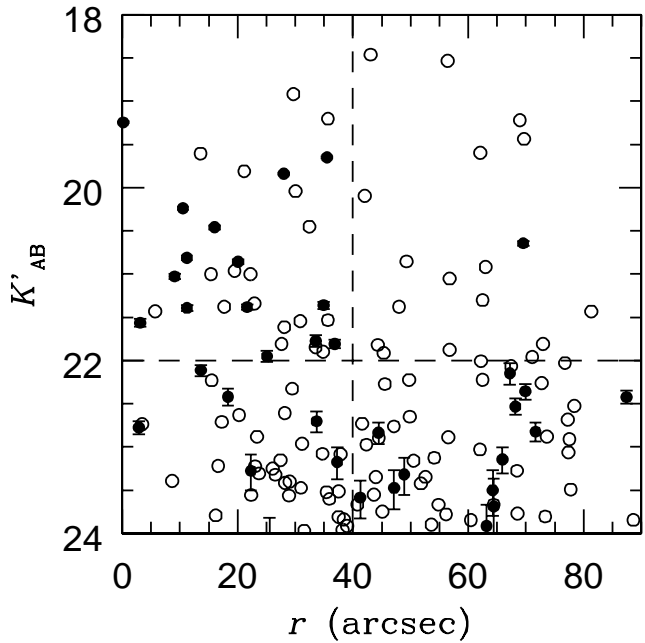}
   \end{center}
   \caption{$K'_{\rm AB}$ magnitude of galaxies against the distance in
   arcsecs from 3C 324. The filled and open circles indicate cluster members
   and non-members, respectively. \label{fig8}}
\end{figure}

Figure 8 plots $K'_{\rm AB}$ magnitude of galaxies against the distance from
3C 324. It is clear that the brighter galaxies are more centrally
concentrated. The filled and open circles indicate the cluster members and
field galaxies, respectively. Note that the distribution of the field galaxies
is quite uniform, which gives indirect support that our photometric redshift
estimation is successful. It is found that most (15/16) of the members with $
K'_{\rm AB}<22$ are located in the inner region. This is in contrast to the
distribution of the faint ($K'_{\rm AB}\geq22$) members. The number (number
density) of the $K'_{\rm AB}\geq22$ galaxies in the inner region and the outer
regions are 6 (4.3 galaxies arcmin$^{-2}$) and 13 (5.0 galaxies arcmin$^{-2}$),
respectively. The significant difference in the distributions of bright and
faint members suggests a strong luminosity segregation of galaxies in the
3C 324 cluster.

\subsection{Luminosity Function}

%
%

\begin{figure}
   \begin{center}
      \FigureFile(85mm,85mm){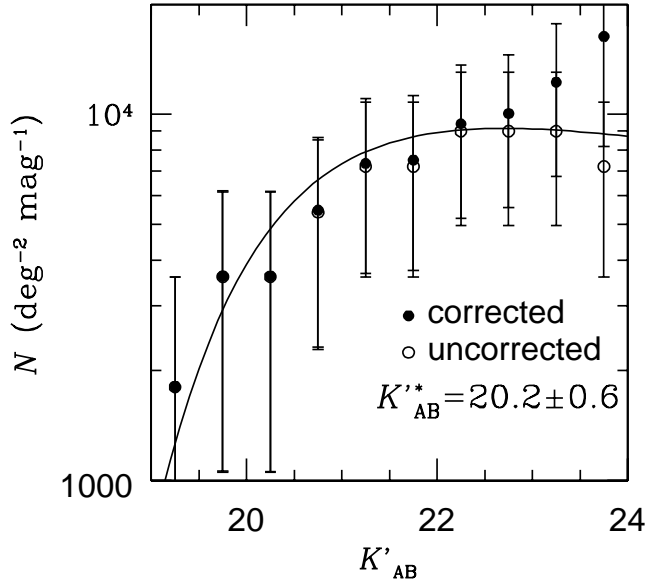}
   \end{center}
   \caption{$K'$-band luminosity function of the cluster galaxies. The open
   and the filled circles indicate the raw counts and the counts corrected for
   incompleteness, respectively. The best-fitted Schechter function is plotted
   as the solid curve. \label{fig9}}
\end{figure}

Figure 9 shows the $K'$-band luminosity function of the cluster galaxies.
The open and filled circles indicate the raw counts and those
corrected for incompleteness, respectively. We fit the Schechter function, $
\phi(L)=\phi^\ast(L/L^\ast)^\alpha\exp(-L/L^\ast)$, to the
completeness-corrected counts in the range 19--23 mag in order to compare the
results with those obtained in a similar manner for lower redshift clusters
by de Propris et al.\ (1999). We fixed the faint-end slope of the Schechter
function to $\alpha=-0.9$ following de Propris et al.\ (1999). We obtained the
characteristic magnitude of $K'^\ast_{\rm AB}=20.2\pm0.6$ mag. The fitted
Schechter function is shown by the solid curve in figure 9. The corrected
3C 324 counts are slightly higher than the fitted Schechter function at the
faint end ($K'_{\rm AB}>23$). These high counts could be
due to dwarf galaxies, because dwarf galaxies start to dominate at $
\sim M^\ast+3$ mag in the $K$-band luminosity function of the nearby Coma
cluster (de Propris et al.\ 1998). However, the uncertainty of the counts at
such faint magnitudes is large.

Our luminosity function was fitted very well by a Schechter form all the way
down to a magnitude of at least $K'_{\rm AB}\sim23$. K00a claimed an abrupt
decrease in their $K'$-band luminosity function of the galaxies in the inner
region at $K'_{\rm AB}\gtsim22$. They obtained the result by subtracting the
galaxy counts in the general field or those in the outer region of the
cluster from the raw counts in the inner region. K00a suggested two possible
interpretations for the abrupt decrease: a strong luminosity segregation
between the inner region and the outer region, or an intrinsic deficiency of
faint galaxies in the 3C 324 cluster as a whole. Now, we find clearly that this
decrease (in the $K'_{\rm AB}\gtsim22$) is due to luminosity segregation.
With the photometric-redshift selected sample, we indeed see several
faint cluster members with $K'_{\rm AB} \gtsim22$, even in the inner region,
which implies that the field subtraction in K00a using the average counts was
also an overestimation, and that the true background counts in this region
seems to be lower than the average of the general field.

%
%

\begin{figure}
   \begin{center}
      \FigureFile(85mm,85mm){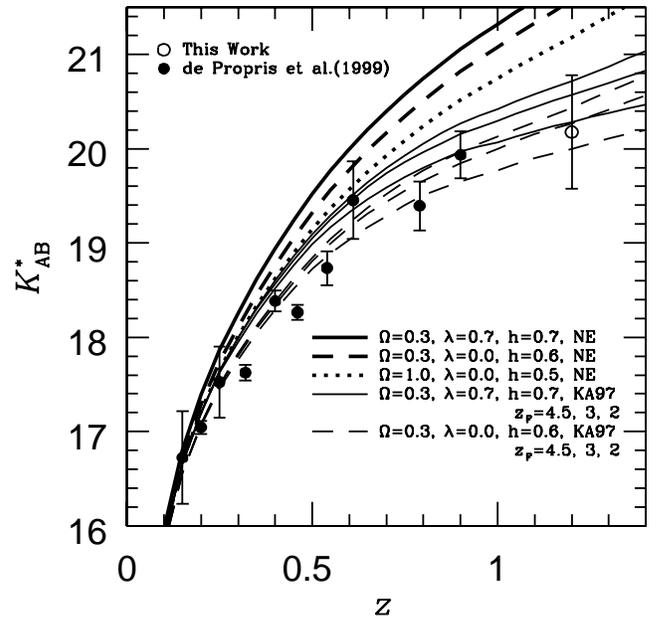}
   \end{center}
   \caption{$K^\ast$--$z$ Hubble diagram for the clusters at $0.1<z<0.9
   $ studied by de Propris et al.\ (1999; filled circles) and the 3C 324
   cluster (open circle). Various lines show the behavior expected for
   no-evolution and passive evolution models with various cosmological
   parameters. $z_{\rm F}$ is the formation redshifts, and NE means no
   evolution. \label{fig10}}
\end{figure}

Figure 10 compares the $K^\ast_{\rm AB}$ value of the 3C 324 cluster with those
of lower-redshift clusters studied by de Propris et al.\ (1999). Note that we
transformed our $K'^\ast_{\rm AB}$ to $K^\ast_{\rm AB}$ in this figure to be
consistent with de Propris et al.\ (1999). Various lines show the expected
loci of the no-evolution and passive evolution models with various
cosmological parameters calculated using Kodama and Arimoto's (1997; KA97)
stellar population synthesis model. It is seen that our result follows the
trend of the intermediate-redshift clusters, which is consistent with the
passive-evolution models. Accordingly, at $K'_{\rm AB} \ltsim K'^\ast_{\rm AB}$
mag, the dominant population in the 3C 324 cluster seems to be old quiescent
galaxies, which were formed at least $\sim1$~Gyr prior to the observed epoch.

\subsection{The Color Properties}

\subsubsection{Color--magnitude diagrams}

%
%

\begin{figure*}
   \begin{center}
      \FigureFile(180mm,85mm){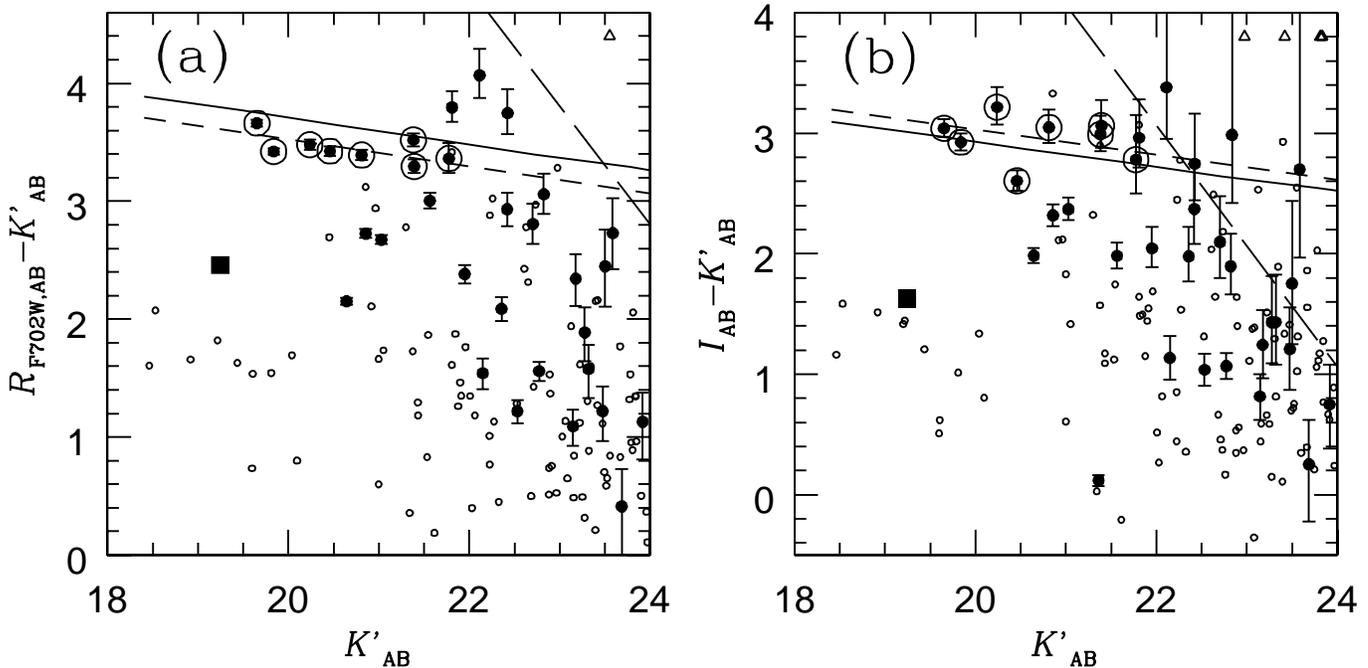}
   \end{center}
   \caption{Color--magnitude diagrams of the cluster members [(a) $
   R_{\rm F702W,AB}-K'_{\rm AB}$ versus $K'_{\rm AB},$ (b) $I_{\rm AB}-
   K'_{\rm AB}$ versus $K'_{\rm AB}$]. The filled circles indicate cluster
   members, and the small open circles indicate non-members. The filled
   square in each panel shows 3C 324, and the long-dashed line in each panel
   shows the 3$\sigma$-detection limit of $R_{\rm F702W,AB}$ and $I_{\rm AB}
   $, respectively. The solid line corresponds to a predicted color--magnitude
   relation of passively evolving galaxies formed at $z_{\rm F}=4.5$. Eight
   bright galaxies, which make a tight color--magnitude relation, are
   indicated by a circles. The short dashed line in each panel shows a fit of
   a straight line to these 8 galaxies with its slope fixed to that of the
   model prediction. \label{fig11}}
\end{figure*}

Figure 11 shows the color--magnitude diagrams of the cluster members [(a) $
R_{\rm F702W, AB}-K'_{\rm AB}$ versus $K'_{\rm AB}$, (b) $I_{\rm AB}-
K'_{\rm AB}$ versus $K'_{\rm AB}$]. The filled circles indicate the cluster
members, and the small open circles indicate non-members. The filled square
in each panel shows 3C 324, and the long-dashed line in each panel shows the
3$\sigma$-detection limit of $R_{\rm F702W,AB}$ and $I_{\rm AB}$,
respectively. The solid line shows a predicted color--magnitude relation at
the cluster redshift ($z=1.2$) for passively evolving galaxies formed
at $z_{\rm F}=4.5$ (KA97). This model assumes that the color--magnitude
relation is a pure metallicity sequence as a function of the galaxy luminosity,
and calibrates the relation by using that of Coma early-type galaxies
(hereafter `Coma C--M model'). In figure 11a, it is found
that bright 8 galaxies (marked with circles) represent a tight color--magnitude
relation (or so-called `red finger'). The short-dashed line in each panel
shows a linear regression line to these red-finger galaxies with the slope
fixed to that of the model prediction. The color differences between the model
line and the fitted line are $-0.183$ mag and 0.103 mag for (a) and (b),
respectively. We find that the scatters of red-finger galaxies around the
best-fitted lines are 0.091 mag and 0.193 mag for (a) and (b), respectively.
The scatter for (a) is comparable with those of the clusters at $z<0.9$ given
in Stanford et al.\ (1998), who estimated the dispersion in the
color--magnitude relation to be as small as 0.1 mag for the morphologically
selected early-type galaxies in clusters. The scatter for (b), however, is
twice as large as that for (a). This may be due to the relatively large
photometric errors of the $I$-band image whose quality is not as good as
the $R_{\rm F702W}$-band image.

Note that the overall color distribution of the entire cluster members is
broad by extending toward bluer colors. Similar trends are also seen in several
clusters at high redshift (Stanford et al.\ 1997; Postman et al.\ 1998;
Tanaka et al.\ 2000; van Dokkum et al.\ 2000). This suggests that there are
more star-forming galaxies in clusters at $z\gtsim 1$ than in clusters at
intermediate redshifts. We discuss this trend again in subsection 5.4.
It is also worth noting that most of the relatively blue galaxies are
faint: $K'_{\rm AB}>K'^\ast_{\rm AB}+1.5$. Therefore, the major star
formation of the bright galaxies ($K'_{\rm AB}\sim K'^\ast_{\rm AB}$) was
over at $z=1.2$, as described in subsection 5.2, while the star formation
activity of the faint galaxies was still considerable in the 3C 324 cluster
at $z\sim1.2$.

K00b extracted early-type member galaxies in the 3C 324 cluster using $
R_{\rm F702W}-K'$ and $B_{\rm F450W}-R_{\rm F702W}$ colors. They regarded
a galaxy of a given $R_{\rm F702W}-K'$ as a member if it is redder in $
B_{\rm F450W}-R_{\rm F702W}$ than a model galaxy having the same $R_{\rm F702W}
-K'$ color. They obtained two main results concerning the
color--magnitude relation of early-type galaxies. Firstly, they found that the
color--magnitude sequence is truncated at $K'_{\rm AB}\sim22$. We also
find the truncation clearly in figure 11a. It is in remarkable contrast
with the sequence of low-redshift clusters, which extend over more than 4
magnitudes (See also K00b). This may suggest that the faint ($
K'_{\rm AB}\gtsim K'^*_{\rm AB}+1.5$) early-type galaxies did not form
untill $z\sim1.2$.

Secondly, K00b found that the early-type galaxies seem to form a broad
color--magnitude sequence whose slope is much steeper than that predicted by
coeval passive-evolution models (see figure 9 of K00b), especially at the faint
end of $K'_{\rm AB} > 22$. They suggested a possible interpretation for this
aspect, that the `tilt' of the slope is due to the age difference along the
sequence; i.e., fainter galaxies are younger than brighter galaxies.
We have confirmed that the slope of the red-finger-galaxies sequence ($
K'_{\rm AB }<22$) is consistent with that predicted by coeval passive-evolution
models (Dickinson 1995; Kodama et al.\ 1998; K00b), but do not clearly confirm
the sequence of red galaxies at $K'_{\rm AB}>22$, which makes the slope
significantly steeper over the whole magnitude range in K00b. Indeed, we
find only three member galaxies on the blue side of the red sequence at $
K'_{\rm AB}=22$--23, where K00b classified more galaxies as cluster members by
their color selection. In this sense, there is no clear age-difference effect
seen in the color--magnitude sequence of red galaxies.

Van Dokkum et al.\ (2001) investigated the color--magnitude relation of the
cluster RXJ 0848+4453 at $z=1.27$. They found that the color--magnitude
relation of early-type galaxies in this cluster is shallower than that for
the nearby Coma cluster, and deduced that the brightest early-type
galaxies may have young stellar populations at $z=1.27$. However, we do not
find such a trend in the 3C 324 cluster. There might be a variation in the
slope of the color--magnitude relation among clusters at $z\sim1.2$.

\subsubsection{Distribution of redder galaxies}

%
%

\begin{figure*}
   \begin{center}
      \FigureFile(180mm,85mm){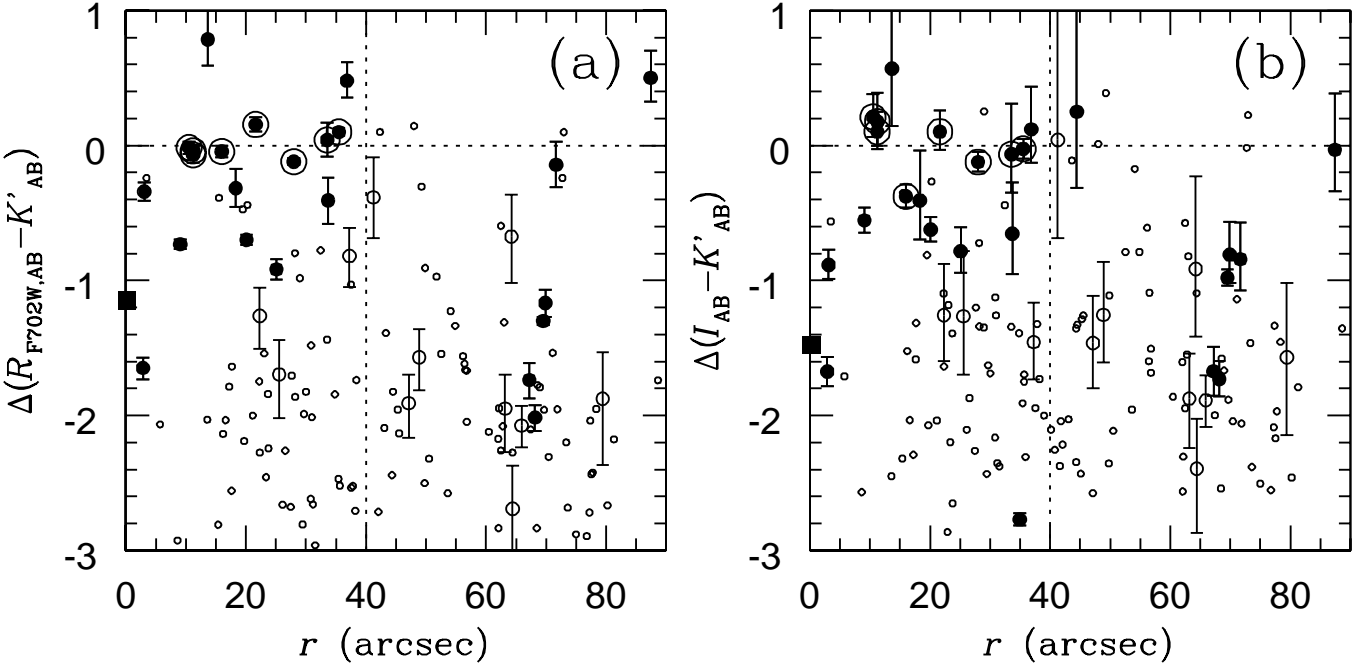}
   \end{center}
   \caption{Relation between the `differential color' and the projected
   distance from 3C 324 for all galaxies, where differential color is defined
   as the observed color minus the color of the red sequence (long-dashed
   lines in figure 11) at the same $K'$ magnitude. The large filled circles
   indicate the cluster members at $K'_{\rm AB}<23$, while the faint cluster
   members at $K'_{\rm AB}>23$ are shown as open circles. The small open
   circles indicate the non-members. The red-finger galaxies defined in
   figure 11 are indicated by circles, and the solid square in each panel
   corresponds to 3C 324. \label{fig12}}
\end{figure*}

Figure 12 shows the color deviation from the red color--magnitude sequence
as a function of the projected distance from 3C 324 for all galaxies.
The large filled circles indicate the cluster members with $K'_{\rm AB}<23
$, while the faint cluster members with $K'_{\rm AB}>23$ are shown as large
open circles (since uncertainties in the photometric redshifts for the faint
galaxies may be large due to their large photometric errors). The small open
circles indicate the non-members. The large circles enclosing the large filled
circles indicate the red-finger galaxies defined in figure 11, and the solid
square in each panel shows the 3C 324 galaxy. We find that the red cluster
members are highly concentrated in the inner region, while the bluer ones are
more widely distributed. We refer to this as `color segregation' analogous to
the `luminosity segregation' found in subsection 5.1. Since the bright and red
galaxies are likely to be early-type galaxies (as in the lower redshift
clusters), the `color segregation' and the `luminosity segregation' may be
reflecting the morphology segregation or morphology-density relation discussed
by Dressler (1980) and Dressler et al.\ (1997) that the early-type galaxies
are concentrated towards the cluster center. The bright early-type galaxies
at $K'_{\rm AB}<22$ ($\sim K'^\ast_{\rm AB}+2$) might have been formed near
the cluster center. Note that the color distribution of the
field galaxies is quite uniform against the distance from 3C 324, similar to
the magnitude distribution shown in figure 8, which again supports our
separation of field galaxies based on the photometric redshifts.

It is also worth pointing out that the physical scale of 40$''$ at $z=1.2$ is
very small (0.33 Mpc). The early-type galaxies in the local clusters are
usually distributed over $\sim1$~Mpc from the cluster center (e.g., Godwin et
al.\ 1983 [Coma cluster]), and those of the MS 1054 cluster at $
z=0.83$ are also distributed over $\sim1$~Mpc (See figure 10 of van Dokkum et
al.\ 2000). This suggests that the 3C 324 cluster is more compact in
size compared with the local clusters, and possibly to the clusters at
intermediate redshifts.

\subsubsection{2-color diagram}

%
%

\begin{figure*}
   \begin{center}
      \FigureFile(130mm,130mm){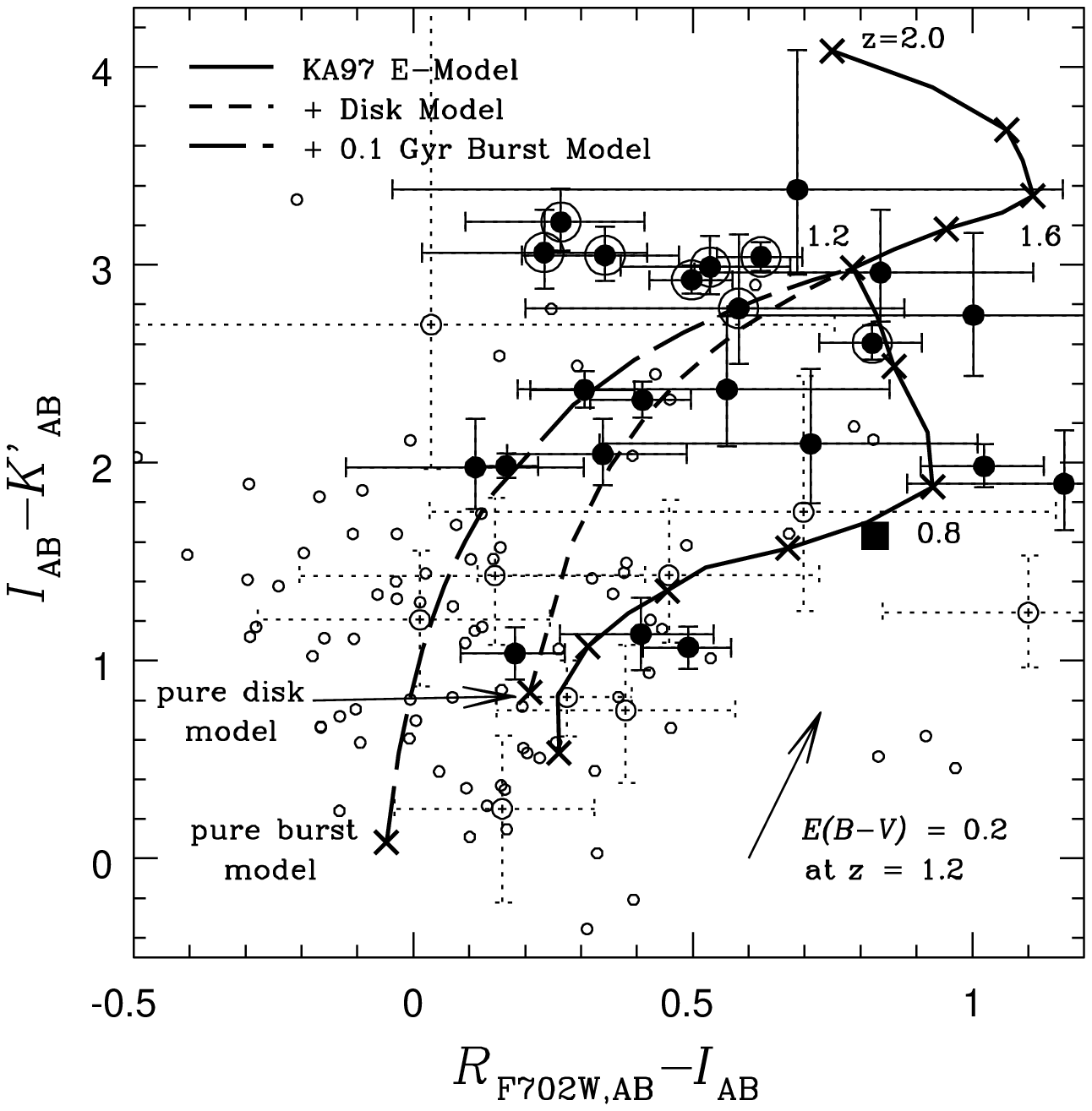}
   \end{center}
   \caption{$R_{\rm F702W,AB}-I_{\rm AB}$ versus $I_{\rm AB}-K'_{\rm AB}$ of
   the objects detected in the 3C324 field. The meaning of symbols is the
   same as in figure 12. The solid line is a prediction of passively evolving
   galaxies at $0\leq z\leq2$ ($M_V=-22$) formed at $z_{\rm F}=4.5$. The disk
   and burst models are also shown as the short-dashed line and the
   long-dashed line, respectively. The effect of internal extinction is shown
   as arrows in the bottom-right corner. See text for details. \label{fig13}}
\end{figure*}

In figure 13, we show an $I_{\rm AB}-K'_{\rm AB}$ versus $R_{\rm F702W,AB}-
I_{\rm AB}$ diagram. The meanings of the symbols are the same as in figure
12. The solid line shows the evolutionary track from $z=0$ to 2 of the
Coma C--M model for $M_V=-22$. The effect of reddening is shown as arrows
in the bottom-right corner, using the extinction curve of Calzetti et
al.\ (2000). We find that only a few cluster galaxies
are consistent with the Coma C--M model at $z=1.2$. The eight red-finger
galaxies (marked with large open circles) have significantly bluer $
(R_{\rm F702,AB}-I_{\rm AB})$ colors than the Coma C--M models.
Since the $(R_{\rm F702,AB}-I_{\rm AB})$ color approximately
corresponds to ($U-B$) color in the rest frame at $z=1.2$, we call these
galaxies `UV-excess galaxies'. Note that these UV-excess galaxies have
optical ($I_{\rm AB}-K'_{\rm AB}$) colors which are consistent with the
Coma C--M model.

Although there are four cluster members which have abnormally red $
R_{\rm F702W,AB}-I_{\rm AB}$ colors ($R_{\rm F702W,AB}-I_{\rm AB}>0.9$),
we think that these colors are likely to be due to photometric errors. Two of
them are located near to the edges of the images, where the image
quality is worse than in the other part, especially in the $I$-band images
where the fringe patterns could not be completely removed. The remaining two
galaxies are located near a bright star or the 3C 324 galaxy where the
accurate photometry is difficult.

To understand the nature of the UV-excess galaxies, we considered the disk
models and the burst models, as follows. We added a disk or a star-burst onto
the passively evolving galaxy (bulge component). The rest-frame near-UV
colors ($R_{\rm F702W}-I$ at $z\sim1.2$) of these model galaxies were
determined by the ratio of the disk or the star-burst component to the bulge
component. As the disk component, we adopted a continuous star-formation model
with an exponentially decaying time scale of 5~Gyr. As the star-burst
component, we adopted a constant star-formation rate (SFR) model observed
at 0.1~Gyr after the onset of star formation. Both of these models were
calculated using the population synthesis code of KA97 with the Salpeter
IMF (mass cutoffs $M_{\rm l}=0.1\MO,$ and $M_{\rm u}=60\MO$). For convenience,
we introduced the disk and burst ratios, $f_{\rm disk}$ and $f_{\rm burst}$,
respectively, which are defined here as the flux ratio of the disk or burst
component to the whole galaxy in the rest-frame $B$ band. We changed $f_{\rm
disk}$ and $f_{\rm burst}$ in the range of 0 and 1.

In figure 13, the disk and burst models are shown as the short-dashed line
and the long-dashed line, respectively. Most of the member galaxies at $
I_{\rm AB}-K'_{\rm AB}<2.6$, which are probably late-type galaxies, follow
the model track from the pure passively-evolving model to the pure disk or
the star-burst model. Thus, the broad color distribution of the galaxies in
the cluster region can be explained by a variation of the fraction of the
disk or star-burst component in galaxies. The red galaxies at $I_{\rm AB}-
K'_{\rm AB}>2.6$, however, cannot be explained by these model tracks. It is
worth noting that the fraction of such objects is very high in this cluster.
Among the galaxies with $K'_{\rm AB}<23$ and $I_{\rm AB}-K'_{\rm AB}>2.6$, $
\sim60\%$ show UV excess. The mean of $R_{\rm F702W,AB}-I_{\rm AB}$
color of these UV-excess galaxies is $\sim0.35$ mag bluer than that of the
passively evolving galaxy with a similar $I_{\rm AB}-K'_{\rm AB}$ ($\sim3.0
$), and this difference is much larger than the photometric errors in $
R_{\rm F702W}$ and $I$ magnitudes. It is also worth noting that similar
results were obtained by Tanaka et al.\ (2000) for a $z\sim1.1$ cluster and by
Haines et al.\ (2001) for a $z\sim1.2$ cluster.

The existence `UV-excess galaxies' suggests that a residual star formation
might occur in the early-type galaxies at $z\sim1.2$, although the
possibility of a high star forming activity is ruled out in
subsection 5.2 based
on the evolution of $K^\ast_{\rm AB}$. However, we find in the following that
even the burst models cannot explain their blue $U-V$ colors. Tanaka et
al.\ (2000) suggested two possibilities for interpreting these UV-excess
galaxies: an intermittent star formation or the last stage of the tail of
star formation. To examine these possibilities, we computed another burst model
for which a star-burst component has a flatter IMF ($x=0.35$) than the Salpeter
IMF ($x=1.35$). We find that differences in $R_{\rm F702W,AB}-I_{\rm AB}$ are
less than 0.1 mag between the $x=0.35$ model and the Salpeter IMF model,
implying that it is difficult to explain the UV-excess galaxies by flat IMF
models. If we want to explain these UV-excess galaxies with an additional
star formation in the red galaxies, we must introduce a very exotic shape of
IMF, which brightens only UV light and with almost no effect on optical light.
It may seem possible, on the other hand, that the burst or disk models with
dust extinction can explain UV-excess galaxies. However, in order to explain
the tight sequence of the UV-excess galaxies seen in figure 13, we need fine
tuning between $f_{\rm burst}$ or $f_{\rm disk}$ and $E(B-V)$. We have
not identified what causes the observed UV excess.

\subsection{Blue Galaxy Fraction}

There have been few attempts to extend measurements of the Butcher--Oemler
effect (Butcher, Oemler\ 1978, 1984) beyond $z\sim0.5$, primarily because
the field contamination becomes severer at higher redshifts.
We try to extend the study on the Butcher--Oemler effect to this very high
redshift cluster at $z=1.2$ by reducing the field contamination using the
photometric-redshift technique.

Butcher and Oemler (1984) defined the fraction of blue galaxies,
$f_{\rm B}$, which are bluer than the color--magnitude sequence of early-type
galaxies by more than 0.2 magnitude in the rest-frame $B-V$ color.
They imposed a radial cut, $R<R_{30}$, the radius which contains
30\% of the cluster population, and the magnitude cut,
$M_V<-20$ in the rest-frame. Note that they did not take into account the
luminosity and color evolution of galaxies in the above definition of the
blue galaxy fraction.

If we ignore luminosity evolution, the magnitude limit that is equivalent to
the Butcher--Oemler limit of $M_V=-20$ in the rest frame is $
K'_{\rm AB}=22.7$ for the 3C 324 cluster, and a difference of $\Delta(B-V)=0.2
$ mag in the rest frame corresponds to $\Delta(I_{\rm AB}-K'_{\rm AB})\simeq
0.65$ in the observed frame at $z=1.2$. Thus, we define $f_{\rm B}$ of the
3C 324 cluster to be the fraction of galaxies brighter than $K'_{\rm AB}=22.7
$ and bluer in $I_{\rm AB}-K'_{\rm AB}$ than the red-finger
sequence (long-dashed line in figure 11) by more than 0.65 mag in the whole
region observed. The obtained value of $f_{\rm B}$ is $0.39\pm0.28$.
We derived $f_{\rm B}$ using $I_{\rm AB}-K'_{\rm AB}$, not $R_{\rm F702W,AB}-
K'_{\rm AB}$, because at $z=1.2$ the $I$ band is nearer to the $B$ band in
the rest frame than the $R$ band and the $R$-band image taken with the HST
does not cover the whole $K'$-band image.

However, ignoring the luminosity evolution may not be a good assumption at
such a high redshift as $z=1.2$. Indeed, if passive evolution is adopted, $
M_V=-20$ at $z=0$ corresponds to $K'_{\rm AB}=21.7$ at $z=1.2$, which
is one magnitude brighter than that for no evolution. Therefore, we must derive
the value of $f_{\rm B}$ while taking the luminosity evolution of the $M_V
$ and $B-V$ color into account. To estimate the effect of evolution, we
used a model of passively evolving galaxies. A difference of $\Delta(B-V)=0.2
$ mag corresponds to $\Delta(I_{\rm AB}-K'_{\rm AB})=0.39$ at $z=1.2$ if
evolution is considered, and we find $f_{\rm B}$ to be $0.46\pm0.34$. This
value coincides with the value for no evolution within error bars, i.e., the
effect of the evolution on $f_{\rm B}$ is found to be small.

%
%

\begin{figure}
   \begin{center}
      \FigureFile(85mm,85mm){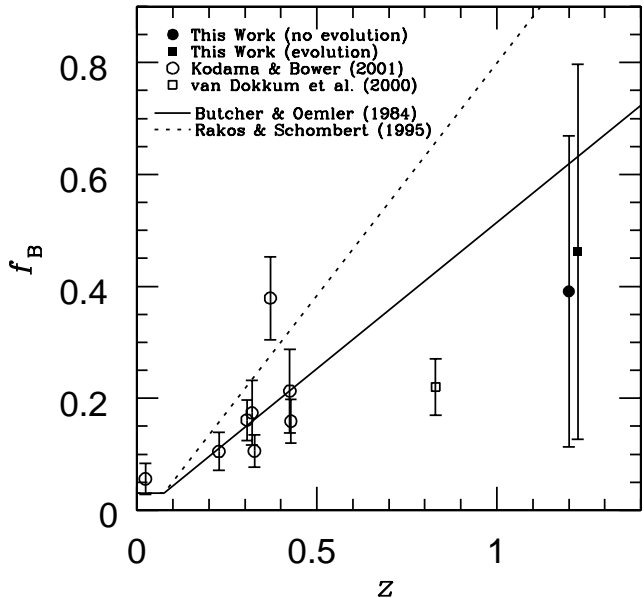}
   \end{center}
   \caption{Blue galaxy fraction $f_{\rm B}$ in rich clusters as a function
   of redshift. The filled square and circle show the result of this work with
   and without evolutionary correction, respectively. The open circles indicate
   7 CNOC clusters and the Coma cluster, and the open square is for MS 1054-03
   at $z=0.83$. The two lines are extrapolations to higher redshifts of the
   observed evolution of $f_{\rm B}$ obtained by Butcher and Oemler
   (1984; solid line) and Rakos and Schombert (1995; dotted line).
   \label{fig14}}
\end{figure}

Figure 14 shows the blue fraction ($f_{\rm B}$) of rich clusters at various
redshifts. The filled square and the filled circle represent our results with
and without an evolutionary correction, respectively. The open circles indicate
7 CNOC clusters and the Coma cluster taken from Kodama and Bower (2001), and
the open square is for MS 1054-03 at $z=0.83$ (van Dokkum et al.\ 2000). The
two lines in figure 14 are linear extrapolations to higher redshifts of the
observed evolution of $f_{\rm B}$ obtained by Butcher and Oemler
(1984; $0<z<0.5$; solid line) and Rakos and Schombert (1995; $0<z<1$; dotted
line). Although the $f_{\rm B}$ of the 3C 324 cluster is found to be higher
than those of $z<1$ clusters, the extremely high fraction of blue galaxies
found by Rakos and Schombert (1995) in
clusters at $z>0.5$ ($f_{\rm B}\sim0.8$ at $z\sim1$) is not confirmed in our
study, similar to the results obtained by van Dokkum et al.\ (2000). The $
f_{\rm B}$-value of the 3C 324 cluster is consistent within the error with the
extrapolation of the evolution obtained in Butcher and Oemler (1984). The
observed area ($\sim1$~Mpc $\times$ 1~Mpc) is probably smaller than a circle
of radius $R=R_{30}$ (typically $\sim 1 h^{-1}$~Mpc in the Butcher
and Oemler 1984). This means that the value of $f_{\rm B}$ obtained in this
work should be taken as the lower limit, because redder galaxies are more
concentrated towards the cluster center. However, since the 3C 324 cluster is
found to be compact in comparison with nearby rich clusters, we think $
f_{\rm B}$ does not increase significantly, even if we exactly follow Butcher
and Oemler's (1984) definition.

We find from figure 14 that the blue galaxy fraction, $f_{\rm B}=0.39\pm
0.28$, of the 3C 324 cluster is higher than those of $z<1$ clusters. This
suggests that the star-formation activity of the entire cluster members in
the 3C 324 cluster is high compared with those of the intermediate clusters.

Note that the 3C 324 cluster is not necessarily a typical cluster at $z\gtsim1
$. We have to observe more clusters at $z\gtsim1$ to derive the general
blue fraction at such a high redshift.

\section{Summary}

We have analyzed multi-photometric data of the cluster 3C 324 at $z\simeq
1.2$ and have identified 35 galaxies as the cluster members with a
photometric-redshift technique. We summarize our findings as follows:

\begin{enumerate}

\item The red and luminous galaxies are distributed in a region enclosed
within a circle of 40$''$ (0.33 Mpc) radius from 3C 324, while bluer galaxies
are distributed more or less uniformly over a wider area. This probably
demonstrates morphology segregation. The giant early-type galaxies might
have been formed near the cluster center. The concentration of early-type
galaxies in such a small region of $r=0.33$~Mpc suggests that the 3C 324
cluster is more compact in size compared with local clusters, and possibly
to clusters at intermediate redshifts in which early-type galaxies generally
extend over $\sim1$~Mpc.

\item The luminosity function of the cluster members is well fitted by a
Schechter function, and the derived characteristic magnitude is $K'^
\ast_{\rm AB}=20.2\pm0.6$ mag. We compared this value with $K^\ast_{\rm AB}$ of
the lower-redshift clusters studied by de Propris et al.\ (1999), and we found
that our result, together with those by de Propris et al.\ (1999), is
consistent with the passive evolution models. This suggests that the giant
early-type galaxies ($\gtsim K'^\ast+1$) are old quiescent
galaxies which were formed at least $\sim1$~Gyr prior to the observed epoch.

\item We confirmed the tight color--magnitude sequences of the bright galaxies
with $19.5 < K_{\rm AB} < 22$ in both $R_{\rm F702W,AB}-K'_ {\rm AB}$ versus $
K'_{\rm AB}$ and $I_{\rm AB}-K'_{\rm AB}$ versus $K'_{\rm AB} $ diagrams.
The slope of the sequence is consistent with the passive evolution model.
On the other hand, truncation of the color--magnitude sequence indicated
by K00b is clearly seen at $K'_{\rm AB}\sim22$. It may suggest that the
faint ($K'_{\rm AB}\gtsim K'^\ast_{\rm AB}+1.5$) early-type galaxies are still
in the process of formation at $z\sim1.2$.

\item The cluster contains members with blue colors at faint magnitudes.
This is also observed in several other $z\sim1$ clusters. This suggests that
although the bright galaxies have finished the major star formation, there
are many faint galaxies ($K'_{\rm AB}\gtsim K'^\ast_{\rm AB}+2$) having
star-forming activity at $z\sim1.2$.

\item Most of the galaxies that have red $I_{\rm AB}-K'_{\rm AB}$ colors are
brighter in the rest UV (bluer in $R_{\rm AB}-I_{\rm AB}$) compared with the
passive evolution models. These UV-excess galaxies have also been found in
other $z>1$ clusters. This suggests that cluster ellipticals show some
residual star formation activity at $z>1$.

\item We measured a fraction of blue cluster members following the
definition of Butcher and Oemler (1984), which is $f_{\rm B}=0.39\pm0.28$.
We compare this $f_{\rm B}$ value with those of other low-redshift clusters,
and find that the 3C 324 cluster has a higher $f_{\rm B}$ than $z<1$ clusters.
This suggests that the star-formation activity of the overall cluster members
in this $z\sim1.2$ cluster is high compared with the clusters at intermediate
redshifts.

\end{enumerate}

\bigskip

A part of the data presented here were taken during a test observing run of
the Subaru telescope. We are therefore indebted to all members of the Subaru
Telescope, NAOJ, Japan. We would like to thank the engineering staffs of
Mitsubishi Electric Co. and Fujitu Co. for technical assistance during the
observations. FN, MK, TK, HF, and MO acknowledge Japan Society for the
Promotion of Science for support through its research fellowships for young
scientists. This work was based in part on observations with the NASA/ESA
Hubble Space Telescope, obtained from the data archive at the Space Telescope
Science Institute, U.S.A., which is operated by AURA, Inc.\ under NASA contract
NAS5--26555.

\end{document}